\documentclass[preprint,pra,showpacs,floatfix]{revtex4}
\usepackage{epsfig}
\usepackage{graphicx}
\usepackage{dcolumn}
\usepackage{amssymb,amsmath}

\begin{document}

\newcommand{\Z}{Z_{\rm eff}}
\newcommand{\Zr}{\Z^{(\rm res)}}
\newcommand{\eps}{\varepsilon}
\newcommand{\eb}{\varepsilon _b}

\title{Scattering of a charged particle from a hard cylindrical solenoid: Aharonov-Bohm Effect}

\author{O. Yilmaz}
\email[Electronic address: ]{o.yilmaz@comu.edu.tr}
\affiliation{Physics Department, Canakkale Onsekiz Mart University, Canakkale, Turkey}

\begin{abstract}
The scattering amplitude $f_k(\alpha,\theta)$ of a charged particle from a long hard cylinderical solenoid is derived by solving the time independent Schr\"{o}dinger equation on a double connected plane. It is a summation over the angular momentum quantum number (partial wave summation):$$f_k(\alpha,\theta)=\frac{-1}{\sqrt{2\pi i k}}\sum_ {m=-\infty} ^ \infty e^{im \theta}\frac{2J_{m+\alpha}(ka)}{H_{m+\alpha}^{(1)}(ka)}\,. $$ 
It is shown that only negative mechanical angular momenta, $m+\alpha < 0$, contribute to the amplitude when the radius of the solenoid goes to zero limit ($a\to 0$) without varying the magnetic induction flux $\alpha=\Phi_B/\Phi_0$ (Flux line). Original Aharonov-Bohm result is obtained with this limit.
\end{abstract}


\maketitle
%%%%%%%%%%%%%%%%%%%%%%%%%%%%%%%%%%%%%%%%%%%%%%
\section{Introduction}\label{sec:intro}

In this work, the non-relativistic scattering of a charged particle from a long ideal solenoid with a finite radius $a$ is studied, e.g. look at the references \cite{R83, AALL84}. The question is considered as an unbound boundary value problem on a doubly connected domain. The scalar potential exists only as an infinite barrier on the cylindrical region. However, the magnetic induction flux appears as a phase in the wave function of the particle everywhere.  
 The non-local effect of magnetic flux on the quantum mechanical wave function of a charged particle is physically detectable even though there is no electromagnetic force on it. Before Aharonow and Bohm drew attention about the effect in 1959 in reference \cite{AB59}, other authors had already written the importance of the electromagnetic potentials on electron motion \cite{F39,ES49}. Yet it is mostly known as the Aharonov-Bohm (AB) effect in the literature. Eventually, it has become a text book material. Every modern quantum mechanics book published contains at least a section discussing AB effect, e.g., see \cite{JJS94, SW13}. Despite of the vast publications about AB effect, the monographs \cite{JH97,PT89} and references in there are worth to point out. The effect has also been verified experimentally in 1960 \cite{C60}.
 
The potential function concept is rooted from the classical motion of a system of particles. The Lagrangian function $L$ can be defined natural way if there exists a potential function $V$ to define the force field acting on the system. Let us consider a system with $n$ degrees of freedom ${\bf q}=(q^i)$ ($i=1, \ldots , n$); its motion is determined by a set of ordinary differential equations (Lagrange's equations)
\begin{equation}\label{eq:lagrange}
\frac{d}{dt}\frac{\partial T}{\partial\dot{q}^i}-\frac{\partial T}{\partial q^i}=Q_i,\,\,\,\,\,\, i=1,\ldots ,n\,,
\end{equation}  
where $T$ is kinetic energy, $Q_i$ is generalized force field. Consequently the equation can be written in homogeneous form if there exists a function $V=V(t,{\bf q},\dot{\bf q})$ such that
\begin{align}\label{eq:force}
Q_i&=\frac{d}{dt}\frac{\partial V}{\partial\dot{q}^i}-\frac{\partial V}{\partial q^i}\nonumber\\
&=\frac{\partial^2 V}{\partial\dot{q}^i\partial\dot{q}^j} \ddot{q}^j+\frac{\partial^2 V}{\partial\dot{q}^i\partial q^j} \dot{q}^j+\frac{\partial^2 V}{\partial\dot{q}^i\partial t}-\frac{\partial V}{\partial q^i}\,.
\end{align}
If $Q_i$ is independent of $\ddot{q}^j$, ($j=1,\ldots,n$), then $\frac{\partial^2 V}{\partial\dot{q}^i\partial\dot{q}^j}=0$, ($i,j=1, \ldots ,n$). It implies that $V=-A_i(t,{\bf q}) \dot{q}^i+\varphi(t,{\bf q})$ \cite{FO93}. Therefore the equations of motion (\ref{eq:lagrange}) can be written in terms of the Lagrangian function $L=T-V$ now. $V$ is called extended or velocity dependent potential function. Electromagnetic force field (Lorentz force) is derived from this sort of potential.

The quantum picture comes into play in a standard way after having the Lagrangian function above. The canonical momenta $p_i$, which are conjugate to the generalized coordinates $q^i$, are defined as $p_i\equiv\frac{\partial L}{\partial \dot{q}^i}$. The Hamiltonian function $H(t,{\bf q}, {\bf p})$ is obtained by means of Legendre transformation $H=\dot{q}^ip_i-L( t,{\bf q}, {\bf p})$ if the Hessian does not vanish: ${ \rm Det} \left(\frac{\partial^2 L}{\partial\dot{q}^i\partial\dot{q}^j}\right)\neq 0 $, which is essential to solve $\dot{q}^i$ in terms of $t$, ${\bf q}$, and ${\bf p}$ from the set of equations $p_i=\frac{\partial L}{\partial \dot{q}^i}$, ($i=1,\ldots,n$). The physical meaning of the Hamiltonian function is the fact that it is the total energy of the system. At this point, the fundamental Poisson's brackets ($\{q^i,q^j\}=0$, $\{q^i,p_j\}=\delta_j^i$, and $\{p_i,p_j\}=0$) are used to quantize the classical Hamiltonian system (Dirac's method):
\begin{equation}
\{q^j,p_k\}=\frac{[\hat{q}^j,\hat{p}_k]}{i\hslash}\,,\,\,\,\,\, j,k=1,\ldots,n\, ,
\end{equation}
where $[\hat{q}^j,\hat{p}_k]\equiv \hat{q}^j\hat{p}_k-\hat{p}_k\hat{q}^j$ is the commutation of operators acting on Hilbert space, $\hslash$ is Planck's constant multiplied by $2\pi$, and $i$ is the imaginary unit. In this picture, time $t$ is a parameter carried into Hilbert space. A function obtained in terms of conjugate operators, $\hat{{\bf q}}$ and $\hat{{\bf p}}$, is an another operator acting on the same Hilbert space. For instance, the eigenvalue equation of the time independent Hamiltonian operator, $\hat{H}(\hat{{\bf q}},\hat{{\bf p}})|\psi\rangle=E|\psi\rangle$, is nothing but just the time independent Schr\"{o}dinger equation. Thus, it is seen that the potential function $V$ has more fundamental role in quantum physics than the force field $Q_i$ in Eq.(\ref{eq:lagrange}).

In the rest of the work, the scattering is considered as a boundary value problem for the wave equation in the case of electromagnetic potentials. 
The scattering from a hard cylinder is solved in section \ref{sec:hard}. The particle cannot get into the cylinder due to an infinite potential barrier in the circular region. This is used as a prototype for the presence of existing a solenoid inside the forbidden region for the particle. The Aharonov-Bohm potential is explained in section \ref{sec:abpot}. The scattering amplitude and total cross section are derived in section \ref{sec:abscat} for the solenoid with a finite radius. It is shown that the limit of vanishing magnetic induction flux is reduced to the hard cylinder solution. Finally the limit to the flux line with zero radius (AB original solution) is examined in section \ref{sec:limit}. Then this limit is obtained directly from the scattering amplitude of the current work. Summary of the results and some future concerns are left to the section of conclusion and discussion.   
%%%%%%%%%%%%%%%%%%%%%%%%%%%%%%%%%%%%%%%%%%
\section{Solution of the force-free Schr\"{o}dinger equation in a doubly connected plane}\label{sec:free}

\subsection{Scattering from a hard cylinder} \label{sec:hard}

Our configuration space $D$ is the two dimensional plane $(x,y)$ with a hole of a finite radius $a$ at the origin. $a$ might be any finite real number $ 0 < \epsilon \leqslant a \leqslant R < \infty$. This is equivalent to say that the particle is not allowed to penetrate the cylindrical region at the origin. That is, the probability of finding the particle in this region vanishes at the first glance. It can be realized in the physical point of view that there exits an infinite potential barrier in this region: $V(x,y)= \infty$ in $\sqrt{x^2+y^2} \leqslant a$ and zero anywhere else in the plane. Hence, the energy eigenfunction $u_E(x,y)$ satisfies the time-independent Schr\"{o}dinger equation outside of the circular boundary. 
\begin{equation} \label{eq:sch}
-\frac{\hslash ^2}{2\mu}\left( \frac{\partial^2}{\partial x^2} + \frac{\partial^2}{\partial y^2}\right) u_E+V(x,y)u_E=Eu_E
\end{equation}
In addition $u_E(x,y)=0$ at $x^2+y^2=a^2$ and required asymptotic behaviour at infinity must be applied on $u_E$ for an acceptable physical solution. From the equation $(x/a)^2+(y/a)^2=1$, it is easy to see that a scale transformation defined by $(x',y')=(x/a,y/a)$ reduces the problem around a cylinder with unit radius. It is also canonical transformation in the background classical dynamics because Poisson brackets remain invariant. Thus, the energy (eigenvalue) is scaled according to $E'=a^2 E$. Note that all these might be understood as a consequence of the Riemann mapping theorem in complex plane. Therefore, knowing the solution to the equation (\ref{eq:sch}) outside of a unit circle provides the solution for a circle with arbitrary radius $a$. However, the coordinates $(x',y')$ are not dimensional quantities any more, but rather real numbers without unit in the complex plane.  

The polar coordinate system $(r' \geqslant 0 ,\,\, 0 \leqslant \theta < 2\pi)$ is clearly the most appropriate one for the present issue: $x'=r' \cos \theta$, $y'=r' \sin \theta$. The  equation (\ref{eq:sch}) and boundary condition become 
 \begin{align}\label{eq:schpol}
\frac{\partial^2u }{\partial r'^2}+ \frac{1}{r'} \frac{ \partial u }{ \partial r' }+ \frac{1}{r'^2}\frac{\partial^2u}{\partial \theta^2}+k'^2u=0, \,\,\mathrm{in}\,\, D:r' > 1
 \\ \label{eq:bc}
u(r'=1,\theta)=0 \,\, \mathrm{at\,\, boundary}\,\, \partial D
 \end{align} 
in polar coordinates, where $k'^2=2\mu Ea^2/ \hslash ^2=a^2k^2 $. Applying the standard method of separation of variable, the solution may be written in the form of $u(r',\theta)=f_m(r')e^{im\theta}$, where $m=0,\pm 1, \pm 2, \ldots$ and $f_m(r)$ is the solution of the radial differential equation
\begin{equation}
z^2\frac{  d^2f_m }{d z^2}+ z \frac{ d f_m }{ d z }+(z^2- m^2)f_m=0,
\end{equation}   
where the radial coordinate is scaled with $z=k'r'$ to get the usual form of the Bessel equation. It has regular singularity at $z=0$ and irregular singularity at $z=\infty$. From the two independent solutions $J_m(kr)$ and 
$N_m(kr)$ (Hereafter $'$ is dropped for simplicity from the scaled variables), known as Bessel functions of first kind and second kind respectively, the series representation of the solution of Eq.(\ref{eq:schpol}) with the boundary condition (\ref{eq:bc}) can be written due to the superposition principle:
\begin{equation} \label{eq:sol}
u(r,\theta)=\sum_ {m=-\infty} ^ \infty c_m e^{im \theta} [J_m(k)N_m(kr)-N_m(k)J_m(kr) ],
\end{equation}
where $c_m$ are the expansion coefficients to be determined from the asymptotic behaviour of the wave function $u(r,\theta)$ as $r\rightarrow \infty$ . In two dimensional scattering problem, the asymptotic behaviour is the following form for a short range potential: 
\begin{equation} \label{eq:asym}
u(r \rightarrow \infty, \theta)=C \left( e^{ikr \cos \theta}+f_k( \theta) \frac{e^{ikr}}{ \sqrt{r} } \right)
\end{equation}
The first term shows the plane wave directed from the left to the impenetrable circle. $C$ is just a normalization factor. $f_k(\theta)$ is called scattering amplitude, which depends only $\theta$. In order to complete the solution, $c_m$ coefficients should be found from Eq.(\ref{eq:sol}) and Eq.(\ref{eq:asym}). First, using the well known asymptotic behaviour of the Bessel functions at large distances,
 \begin{align}\label{eq:first}
J_m(kr)\sim \sqrt{ \frac{2}{ \pi kr } } \cos \left(kr-(m+\frac{1}{2}) \frac{\pi}{2} \right)
 \\ \label{eq:second}
N_m(kr)\sim \sqrt{ \frac{2}{ \pi kr } } \sin \left(kr-(m+\frac{1}{2}) \frac{\pi}{2} \right)
 \end{align} 
$u(r,\theta)$ can be rewritten from Eq.(\ref{eq:sol}) about the same asymptotic region,
\begin{align}\label{eq:form1}
u(r,\theta)\sim\frac{e^{-ikr}}{\sqrt{2\pi kr}}\sum_ {m=-\infty} ^ \infty c_m e^{im \theta}iH_m^{(1)}(k) e^{i\varphi_m} 
+\frac{e^{ikr}}{\sqrt{2\pi kr}}\sum_ {m=-\infty} ^ \infty c_m e^{im \theta}(-i)H_m^{(2)}(k) e^{-i\varphi_m} ,
\end{align}
where a phase $\varphi_m=(m+\frac{1}{2}) \frac{\pi}{2}$ is defined for convenience, $H_m^{(1)}(k)=J_m(k)+iN_m(k)$ and $H_m^{(2)}(k)=J_m(k)-iN_m(k)$ are Hankel functions. On the other hand, the same approximation might be obtained in another form from the Eq.(\ref{eq:asym}) as follows
\begin{align}\label{eq:form2}
u(r,\theta)\sim C \frac{e^{-ikr}}{\sqrt{2\pi kr}}\sum_ {m=-\infty} ^ \infty i^m e^{i(m \theta+\varphi_m)} 
+C\frac{e^{ikr}}{\sqrt{2\pi kr}} \left[ \sqrt{2\pi k} f_k( \theta)+ \sum_ {m=-\infty} ^ \infty i^m e^{i(m \theta-\varphi_m)} \right]\,.
\end{align}
The plane wave have been represented by Jacobi-Anger series, $e^{ikr \cos \theta}=\sum_ {-\infty} ^ \infty i^m e^{im \theta}J_m(kr) $ to derive the asymptotic of the wave function (\ref{eq:form2}) above. Comparing these two series (Eqs. (\ref{eq:form1}) and (\ref{eq:form2})) at arbitrary radial distance $r$, the coefficients of the functions $e^{\pm ikr}$ must be equal as they are linearly independent functions. Consequently, the expansion coefficients are extracted as $c_m=Ci^m/iH_m^{(1)}(k)$. Substituting obtained $c_m$ into Eq. (\ref{eq:sol}) completes the solution of the scattering of a particle from the hard cylinder. Finally, it is easy to find the scattering amplitude $f_k(\theta)$ by writing the wave function $u(r,\theta)$ with the found $c_m$ as follows
\begin{equation} \label{eq:finalsol}
u(r, \theta)=C \left( e^{ikr \cos \theta}-\sum_ {m=-\infty} ^ \infty i^m \frac{J_m(k)}{H_m^{(1)}(k)} e^{im \theta}H_m^{(1)}(kr)  \right)
\end{equation}
Note that the boundary condition at $r=1$ is satisfied clearly. The required asymptotic behaviour of $u(r,\theta)$ is also guaranteed by the Hankel function because its form at large distances is given by
\begin{equation}\label{eq:han1}
H_m^{(1)}(kr)\sim \sqrt{\frac{2}{\pi kr}}e^{i(kr-\varphi_m)}
\end{equation}
Thus, the scattering amplitude defined in Eq. (\ref{eq:asym}) can be found by means of Eqs. (\ref{eq:finalsol}) and (\ref{eq:han1}):
\begin{equation}\label{eq:sampli}
f_k(\theta)= -\sqrt{\frac{2}{\pi k}}\sum_ {m=-\infty} ^ \infty i^m \frac{J_m(k)}{H_m^{(1)}(k)} e^{i(m \theta-\varphi_m)}
\end{equation}
Integration of the differential cross section $d\sigma/d\theta=|f_k(\theta)|^2$ over $\theta$ gives the total cross section. It can be presented by a real phase parameter defined by $\sin \delta_m = |J_m(k)|/\sqrt{(J_m(k))^2+(N_m(k))^2}$. Because Bessel functions defined on real axis are real valued functions, therefore, we have total cross section as
\begin{equation} \label{eq:htcs}
\sigma=\frac{4}{k}\sum_{m=-\infty}^\infty \sin^2 \delta_m
\end{equation}
In the next section, electromagnetic field source (solenoid) will be added to the hard cylinder scattering. The infinitely long solenoid carrying steady electric current inside the cylinder create a magnetic field only inside the cylinder.
%%%%%%%%%%%%%%%%%%%%%%%%%%%%%%%%%%%%%%%%%%%%%%%%%%%%%%%%%%
\subsection{Aharonov-Bohm potential}\label{sec:abpot} 
 
Once having an impenetrable cylinder, we may place a source of field inside. In this work, we consider the current density on the surface of the cylinder such as a solenoid. According to Maxwell's equations, time independent (steady) current density will create time independent magnetic field around the source by means of Biot-Savart law. An infinitely long ideal solenoid will create a constant magnetic induction vector field inside the cylinder along the $z-$axis ${\bf B}=(0,0,B)$ but ${\bf B}=(0,0,0)$ outside, where $B$ is a constant. In general, electromagnetic fields that satisfy the Maxwell's equations can be realized by scalar and vector potentials $(\phi,{\bf A})$.
\begin{equation}\label{eq:fields}
{\bf E}= -\nabla \phi -\frac{1}{c}\frac{\partial{\bf A}}{\partial t},\,\,\,\,\,\,\,\,\,\,  {\bf B}=\nabla\times {\bf A}
\end{equation}
These potential functions are not unique since another set of potentials can be found without changing the fields ${\bf E}$, ${\bf B}$ by gauge transformations:
\begin{equation}\label{eq:gauge}
\tilde{\phi}= \phi -\frac{1}{c}\frac{\partial\chi}{\partial t},\,\,\,\,\,\,\,\,\,\,  \tilde{{\bf A}} ={\bf A}+\nabla\chi,
\end{equation}
where $\chi(t,{\bf x})$ is arbitrary real valued space-time function. 
Aharonov-Bohm potential is known as a vector potential of a solenoid carrying steady electric current $I$. Therefore, the vector potential is time independent vector valued function ${\bf A}({\bf x})$.
\begin{align}\label{eq:abpot}
{\bf A}(x,y)&=\frac{B}{2}(-y,x,0)\,,& \,\,\mathrm{for}\,\,r \leqslant a \nonumber \\
&=\frac{B}{2} \frac{a^2}{x^2+y^2}(-y,x,0)\,,& \,\,\mathrm{for}\,\,r \geqslant a
 \end{align} 
From Eq. (\ref{eq:fields}), the fields ${\bf E}$, ${\bf B}$ are decoupled since there is no contribution to the electric field ${\bf E}$ from the vector potential, $\partial{\bf A}({\bf x})/\partial t=0$. A line integral of the vector potential along a closed curve around the solenoid equals the magnetic flux due to the constant magnetic induction inside the solenoid. This integral $$\oint_\Gamma{\bf A} \cdot d{\bf x}=B\pi a^2=\Phi_B$$ is independent of the shape of the simple loop $\Gamma$ around the solenoid so it is known as a topological property of the doubly connected space.

Moreover, the scalar potential $\phi$ can be considered the hard cylinder potential which prevents the charged particle going inside the solenoid. The scalar potential energy discussed in section \ref{sec:hard} can be written as $V(x,y)=q\phi(x,y)$, where $q$ is the electric charge of the particle (e.g. $q=-e$ is the charge of an electron). Note that $V$ is infinite inside the solenoid, so is $\phi$. Yet this can be assumed a limit of a potential as $V_0\rightarrow \infty$
\begin{align}\label{eq:spot}
V(x,y)&=q\phi=V_0\,,& \,\,\mathrm{for}\,\,r \leqslant a \nonumber \\
      &=0\,,& \,\,\mathrm{for}\,\,r > a.
\end{align} 
Although it is a fictitious potential since we have zero electric field everywhere except on the surface of the cylinder, it might be considered idealized version of a more realistic experimental set up.
In addition, it is not hard to see that the vector potential outside the solenoid may be obtained by a time-independent gauge transformation from ${\bf A}=0$ (natural gauge). Because curl of ${\bf A}(x,y)$ for $r\geqslant a$ vanishes, $ \nabla \times {\bf A}=0$. Therefore we can write $ {\bf A}=\nabla \chi(x,y)$ in the region where we want to solve Schr\"{o}dinger wave equation. Here $\chi(x,y)= -(Ba^2/2)\tan^{-1}(y/x)$ is the phase function of the gauge transformation. It is important to note that the phase $\chi$ is not a single valued function, so there is a branch cut on the negative $x-$axis. However, it is not possible to find such a gauge transformation for the vector potential inside the solenoid, where is out of our domain.

\subsection{Scattering from impenetrable long solenoid}\label{sec:abscat}

In the Lagrangian formalism, the interaction potential $V$ between a charged particle and electromagnetic fields is a velocity dependent potential, namely $V=q\phi-\frac{q}{c}{\bf A} \cdot {\bf v}$ (see section \ref{sec:intro}). Here, ${\bf v}=d{\bf x}/dt$ is the velocity vector of the particle. Hence the Lagrangian $L=\mu{\bf v}^2/2-V$ is usually used to define the conjugate canonical momentum ${\bf p}$ to the position vector ${\bf x}$ such that the required relations of the fundamental Poisson brackets are satisfied, $\lbrace x_i,p_j\rbrace=\delta_{ij}$. In the presence of the vector potential, the canonical momentum ${\bf p}$ and kinematical momentum ${\bf \Pi}=\mu{\bf v}$ do not equal to each other, rather they have the following relation
\begin{equation}
{\bf p}=\frac{\partial L}{\partial {\bf v}}={\bf \Pi}+\frac{q}{c}{\bf A}.
\end{equation}  
Having the canonical pair $(x_j,p_j)$ for the above Lagrangian, we know the Hamiltonian for the particle in electromagnetic fields: $H={\bf \Pi}^2/2\mu+q\phi$. 
It can be interpreted physically that scalar potential contributes to the particle energy via its electrical charge while the vector potential via its mechanical (kinematical) momentum. The canonical momentum ${\bf p}$ has a physical meaning in only the case of absence of the vector potential. Its combination with the vector potential, ${\bf \Pi}={\bf p}-\frac{q}{c}{\bf A}$, can determine the real trajectory of the particle classically. Under gauge transformation, Eq.(\ref{eq:gauge}), the dynamics of the particle does not chance since the fields ${\bf E}$, ${\bf B}$ remain the same and the position ${\bf x}$ and ${\bf \Pi}=\mu {\bf v}$ are determined by them. Note that the canonical momentum ${\bf p}$ is obviously not a gauge invariant quantity. The calculation of the classical dynamics may be performed by well-known Hamilton-Jacobi (HJ) partial differential equation satisfied by the Hamilton principle function $S(t,{\bf x},{\bf P})$ as well. $S$ is a generating function of a special canonical transformation and the constant ${\bf P}$ vector is the new canonical momentum after integration of the HJ equation. This is the closest interpretation to quantum mechanical description of the dynamics. The gauge invariance is guaranteed in this interpretation as long as the transformed Hamilton principle function $\tilde{S}$ is given by $\tilde{S}=S+\frac{q}{c}\chi$ under gauge transformation defined in Eq.(\ref{eq:gauge}). 

In the light of the explanation above, the Hamiltonian of a particle with charge $q$ and mass $\mu$ outside the solenoid becomes $H={\bf \Pi}^2/2\mu$ as the scalar potential vanishes in the domain of the problem. Physically, this is a free particle Hamiltonian equivalent to the one in section \ref{sec:hard}, where mechanical momentum consists of only canonical momentum. On the other hand, quantization must be performed according to usual canonical quantization prescription: ${\bf p}$ in classical Hamiltonian is replaced by the operator $-i\hslash\nabla$ in the position representation. Therefore time independent wave equation (Schr\"{o}dinger equation) has the form of $\frac{1}{2\mu}\left(\frac{\hslash}{i}\nabla -\frac{q}{c}{\bf A}\right)  ^2u_E=Eu_E$
%\begin{equation} \label{eq:absch}
%\frac{1}{2\mu}\left(\frac{\hslash}{i}\nabla -\frac{q}{c}{\bf A}\right)  %^2u_E=Eu_E
%\end{equation}
outside the solenoid. Substituting the vector potential from Eq. (\ref{eq:abpot}) explicitly in this equation gives
\begin{equation} \label{eq:absch}
-\frac{\hslash ^2}{2\mu}\left(\frac{\partial^2}{\partial x^2} + \frac{\partial^2}{\partial y^2}+\frac{2i\alpha\left(x\frac{\partial}{\partial y} - y\frac{\partial}{\partial x}\right)- \alpha^2}{x^2+y^2}\right) u_E=Eu_E
\end{equation}
with the boundary condition on the surface of the solenoid; $u_E(x,y)=0$ at $x^2+y^2=a^2$. Here $\alpha=\Phi_B/\Phi_0$ is the dimensionless magnetic flux with the unit of $\Phi_0=-2\pi\hslash c/q$. As in section \ref{sec:hard}, we can make the scale transformation $(x',y')=(x/a,y/a)$ to map the boundary to the circle with unit radius at the origin. Thus Eq. (\ref{eq:absch}) becomes 
 \begin{align}\label{eq:abschpol}
\frac{\partial^2u }{\partial r'^2}+ \frac{1}{r'} \frac{ \partial u }{ \partial r' }+ \frac{1}{r'^2}\left( \frac{\partial }{\partial \theta}+i\alpha\right)^2 u+k'^2u=0
 \end{align}  
in the polar coordinates. Here $k'=ak$ as before and the eigenfunction $u(r',\theta)$ must vanish on the boundary (at $r'=1$) since the particle is hitting an infinite potential barrier. It is seen that the magnetic flux $\alpha $ appears as some portion of the physical angular momentum $L_z$ about $z-$axis of the particle additional to canonical momentum; $L_z=-i\hslash \frac{\partial}{\partial\theta}-\frac{q\Phi_B}{2\pi c}$. Consequently, the angular momentum has eigenvalues $\hslash(m+\alpha)$ with eigenfunctions $e^{im\theta}$, $m=0,\pm 1,\pm 2,\cdots$. The spectrum consists of continuum part, $\hslash \alpha$, in angular momentum quantum number. After separating the angular part of the eigenfunction $u(r',\theta)$, we have the radial equation from (\ref{eq:abschpol}) as
\begin{equation}\label{eq:abrad}
r^2\frac{  d^2f_m }{d r^2}+ r \frac{ d f_m }{ d r }+(r^2k^2- (m+\alpha)^2)f_m=0\,,
\end{equation}
where we dropped $'$ from $k'$ and $r'$ for convenience.   
Note that the magnetic flux contributes only through the index of the Bessel functions. The radial eigenfunctions $f_m(r)$ are proportional to Bessel functions $J_{m+\alpha}(kr)$ and $N_{m+\alpha}(kr)$ with non-integer index $m+\alpha$.

On the other hand, the solution of the wave function for AB potential should be obtained by a unitary rotation in Hilbert space from the hard cylinder solution in section \ref{sec:hard}. Because AB vector potential in field free region can be obtained by a gauge transformation from the trivial gauge ($\phi=0$, ${\bf A}=0$) of the hard cylinder potential (see section \ref{sec:abpot}) : $ \tilde{\phi}=-\partial \chi/c\partial t $, $\tilde{{\bf A}}=\nabla\chi$. Under this transformation wave function of the trivial gauge should gain a phase to leave the wave equation gauge invariant,
\begin{equation}\label{eq:abwf}
\tilde{u}(r,\theta)=e^{i\frac{q}{\hslash c}\chi(r,\theta)}u(r,\theta)
\end{equation} 
It is easy to find the gauge function $\chi$ from the Eq.(\ref{eq:abpot}). 
It is time independent since the vector potential outside the solenoid is time independent. Therefore there is no change occur on the scalar potential due to the transformation, which is zero outside the solenoid originally,
\begin{equation}
\chi(x,y)=\frac{\Phi_B}{2\pi}\arctan(\frac{y}{x}) 
\end{equation}
It is only depends on $\theta$ in polar coordinates and it is independent from the scale transformation of the coordinates as long as the flux $\Phi_B$ is kept constant. Hence, the wave function of the hard cylinder scattering gets extra phase due to the Eq.(\ref{eq:abwf}), $\tilde{u}=e^{-i\alpha \theta} u$. Since the solution of the hard cylinder $u(r,\theta)$ is constructed as a single valued function the transformed one $\tilde{u}(r, \theta)$ is not single valued since $\alpha$ is not an integer. The origin is a branch point and the negative $x-$axis is a branch cut. We should restrict the function to a single branch, $-\pi<\theta<\pi$, in order to make it single valued. However this is the solution in the presence of the vector potential in additional to hard cylinder barrier. It can be checked in the incoming plane wave region where the solution simply must behave like $e^{i(kr \cos \theta-\alpha\theta)}$ according to the gauge transformation so that the probability current density must be  ${\bf J}=(\hslash k/\mu){\bf \hat{x}}$ physically like in the case of hard cylinder in that asymptotic region. Here ${\bf \hat{x}}$ is the unit vector in $x$ direction. Indeed, it can be derived that it is so by using the current density in the case of presence of vector potential:
\begin{equation}\label{eq:abcd}
{\bf J}=\frac{\hslash}{\mu}(u^*\nabla u-u\nabla u^*+\frac{\alpha}{r} |u|^2 {\bf \hat{\theta}})=\frac{\hslash k}{\mu}{\bf \hat{x}}
\end{equation} 
Thus, the solution to the Eq.(\ref{eq:abschpol}) satisfying the boundary condition at $r=1$,  
\begin{widetext}
\begin{equation}\label{eq:absol}
u(r,\theta)=\sum_ {m=-\infty} ^ \infty c_m e^{im \theta} [J_{m+\alpha}(k)N_{m+\alpha}(kr)-N_{m+\alpha}(k)J_{m+\alpha}(kr)]
\end{equation}
\end{widetext}
must show the same feature like in the asymptotic region as well. Indeed it
represents a holomorphic function of $kr$ ($kr$ is considered as a complex variable) throughout the complex plane cut along the negative $x$-axis because the index of Bessel functions $m+\alpha$ is not an integer \cite{AS64}. However the cut line is naturally excluded from the domain because $kr>0$ for a particle with non negative energy $E>0$. The expansion coefficients $c_m$ are determined from the asymptotic behaviour of the solution as $r\to \infty$. Because of the current density requirement from Eq. (\ref{eq:abcd}) 
the asymptotic behaviour of the solution at infinities should be given by
\begin{equation} \label{eq:abasym}
u(r \rightarrow \infty, \theta)=C \left( e^{i(kr \cos \theta-\alpha\theta)}+f_k(\alpha,\theta) \frac{e^{ikr}}{ \sqrt{r} } \right)\, .
\end{equation}
Series representation of it can be obtained by using the Fourier series representation of the phase contribution: 
\begin{equation}\label{eq:fourier}
e^{-i\alpha\theta}=\sum_ {n=-\infty} ^ \infty \frac{\sin(\alpha+n)\pi }{(\alpha+n)\pi}e^{in\theta},\,\,\,\,\,-\pi<\theta<\pi\, .
\end{equation} 
This enables us to use the Fourier expansion of the function $ e^{i(kr \cos \theta-n\theta)}=\sum_{-\infty}^\infty a_m e^{i m\theta} $, where the expansion coefficients are well-known integral \cite{AS64}
\begin{eqnarray}
a_m = \frac{1}{2\pi}\int_{-\pi}^\pi e^{i(kr \cos \theta-(m+n)\theta)}\,d\theta = i^{m+n}J_{m+n}(kr)\,.
\end{eqnarray}
Therefore, we may have the series representation for the plane wave with additional phase as
\begin{equation}
e^{i(kr \cos \theta-\alpha\theta)}=\sum_ {m,n=-\infty} ^ \infty  i^{m+n} e^{im\theta}\frac{\sin(\alpha-n)\pi }{(\alpha-n)\pi}J_{m+n}(kr) 
\end{equation} 
instead of Jacobi-Anger expression in the hard cylinder case (see section \ref{sec:hard}). Instantly, it is clear that it is reduced Jacobi-Anger formula when $\alpha\to 0$ since only $n=0$ term survives in this limit. Hence it is not difficult to express Eq.(\ref{eq:abasym}) as linear combination of incoming and outgoing circular waves:
\begin{widetext}
\begin{align}\label{eq:abform1}
u(r,\theta)\sim & C \frac{e^{-ikr}}{\sqrt{2\pi kr}}\sum_ {m=-\infty} ^ \infty \sum_ {n=-\infty} ^ \infty  i^{m+n} e^{i(m \theta+\varphi_{m+n})}\frac{\sin(\alpha-n)\pi }{(\alpha-n)\pi} \nonumber \\
+&C\frac{e^{ikr}}{\sqrt{2\pi kr}} \left[ \sqrt{2\pi k} f_k(\alpha , \theta)+ \sum_ {m=-\infty} ^ \infty \sum_ {n=-\infty} ^ \infty  i^{m+n}  e^{i(m \theta-\varphi_{m+n})}\frac{\sin(\alpha-n)\pi }{(\alpha-n)\pi}  \right] \, .
\end{align}
\end{widetext}
The Eq.(\ref{eq:absol}) gives another form for asymptotic expansion of the eigenfunction $u(r,\theta)$ as $r\to \infty$ by means of the Bessel functions asymptotic formulae (\ref{eq:first}, \ref{eq:second}). Thus the second form of the asymptotic expansion of the eigenfunction is
\begin{widetext}
\begin{align}\label{eq:abform2}
u(r,\theta)\sim\frac{e^{-ikr}}{\sqrt{2\pi kr}}\sum_ {m=-\infty} ^ \infty c_m e^{im \theta}iH_{m+\alpha}^{(1)}(k) e^{i\varphi_{m+\alpha}} 
+\frac{e^{ikr}}{\sqrt{2\pi kr}}\sum_ {m=-\infty} ^ \infty c_m e^{im \theta}(-i)H_{m+\alpha}^{(2)}(k) e^{-i\varphi_{m+\alpha}} \, .
\end{align}
\end{widetext}
at large distances from the solenoid. Here the Hankel's functions has non integer index $m+\alpha$ and the phase $ \varphi_{m+n} $ is defined as in section \ref{sec:hard}. Consequently, the unknown expansion coefficients $c_m$ can be extracted by comparing the two forms of the asymptotic expansions of the eigenfunction far away from the solenoid. This is allowed because of the uniqueness of the solution of the eigenvalue equation (\ref{eq:abschpol}). Comparing the coefficients of $e^{-ikr}/\sqrt{r}$ in equations (\ref{eq:abform1}) and (\ref{eq:abform2}) provides
\begin{align}\label{eq:abcm}
c_m=C\frac{i^m}{iH_{m+\alpha}^{(1)}(k)}\sum_ {n=-\infty} ^ \infty i^{n} e^{-i(\alpha-n)\frac{\pi}{2}}\frac{\sin(\alpha-n)\pi }{(\alpha-n)\pi}
=C\frac{i^m}{iH_{m+\alpha}^{(1)}(k)}e^{-i\alpha(\varepsilon-\frac{\pi}{2})}\,,
\end{align}
where the limit $ \varepsilon \to 0$ is assumed in the second equality. Eq.(\ref{eq:fourier}) is needed to obtain the expression (\ref{eq:abcm}). 
Once having the $c_m$s, equating the coefficients of $e^{ikr}/\sqrt{r}$
in Eqs. (\ref{eq:abform1}) and (\ref{eq:abform2}) identifies the scattering amplitude $f_k(\alpha,\theta)$ in the presence of the solenoid in the hard cylinder:
\begin{widetext}
\begin{align}\label{eq:abscatam}
f_k(\alpha,\theta)=&-\frac{1}{\sqrt{2\pi k}}\sum_ {m=-\infty} ^ \infty \sum_ {n=-\infty} ^ \infty  e^{i(m \theta-\varphi_{\alpha-n})}\frac{\sin(\alpha-n)\pi }{(\alpha-n)\pi}
\frac{H_{m+\alpha}^{(2)}(k)e^{-i(\alpha-n)\frac{\pi}{2}}+H_{m+\alpha}^{(1)}(k)e^{i(\alpha-n)\frac{\pi}{2}}}{H_{m+\alpha}^{(1)}(k)}\nonumber \\
=&-\frac{e^{-i\alpha\frac{ \varepsilon }{2}} }{\sqrt{2i\pi k}}\sum_ {m=-\infty} ^ \infty e^{im \theta}\frac{e^{i\alpha\frac{ \varepsilon }{2}}H_{m+\alpha}^{(1)}(k)+e^{-i\alpha\frac{ \varepsilon }{2}}H_{m+\alpha}^{(2)}(k)}{H_{m+\alpha}^{(1)}(k)}\,,
\end{align}
\end{widetext}
where it should be understood that the limit $ \varepsilon \to 0$ must be taken in the final expression of differential cross section, $d\sigma/d\theta=|f_k(\alpha,\theta)|^2$. It should be reduced to the scattering amplitude of the hard cylinder without magnetic flux in the limit $\alpha\to 0$. It can be inferred  from the first line of Eq.(\ref{eq:abscatam}) that only contribution to the double series comes for $n=0$ as $\alpha\to 0$ since $\sin n\pi=0$ vanishes for any non-zero integer $n$. Consequently, we first take $n=0$ in the double sum then $\sin \alpha\pi/\alpha\pi\to 1$ in the limit $\alpha\to 0$ gives the the formula (\ref{eq:sampli}). On the other hand, in reference \cite{AB59} the scattering amplitude is derived for the solenoid with zero radius $a=0$. Our solution (\ref{eq:abscatam}) is 
for any finite radius of the solenoid because the coordinates have been transformed to have a solenoid with unit radius by means of a simple scale transformation. Thus we can take the limit of $a\to 0$ in Eq.(\ref{eq:abscatam}) only after we undo the scale transformation. 
As a conclusion, the angular momentum summation for the total cross section becomes
\begin{equation}\label{eq:abtcs}
\sigma=\frac{4}{k}\sum_{m=-\infty}^\infty \frac{|J_{m+\alpha}(k)|^2}{(J_{m+\alpha}(k))^2+(N_{m+\alpha}(k))^2}\,,
\end{equation}
which reduces to Eq.(\ref{eq:htcs}) of the hard cylinder scattering when the flux vanishes ($\alpha\to 0$). If $\alpha$ is expressed as $\alpha=[\alpha]+\nu$, where $[\alpha]$ is the integer part of it, only decimal part $\nu$ make (\ref{eq:abtcs}) distinguish from the total cross section of the hard cylinder, (\ref{eq:htcs}).
%%%%%%%%%%%%%%%%%%%%%%%%%%%%%%%%%%%%%%%%%%%%%%%%%%%%%%%%%%%%%%%%%%%%%%%%
\subsection{Aharonov-Bohm solution as a limit case}\label{sec:limit}

In this section, we would like to make a second test whether the present method can reproduce the AB solution \cite{AB59} when the radius vanishes. As mentioned in the previous section, we need the solenoid radius $a$ as a parameter in the solutions in order to take the limit $a\to 0$. Hence the scale transformation made in section \ref{sec:abscat} is abandoned here, i.e., in any scaled solution, $k'=ka$ and $k'r'=kr$ are replaced with the unscaled ones. It is well known that the solution of the Schr\"{o}dinger equation as $a\to 0$ is reduced to \cite{AB59}
\begin{equation}\label{eq:zero}
u(r,\theta)=\sum_ {m=-\infty} ^ \infty c_m e^{im \theta} J_{| m+\alpha |}(kr)\, .
\end{equation} 
Following the same steps as in sections \ref{sec:hard} and \ref{sec:abscat}, it is easy to obtain the expansion coefficients $c_m$ from the asymptotic form of the eigenfunction; see Eqs. (\ref{eq:abasym}) and (\ref{eq:abform1}). Yet they must be compared the asymptotic form of the solution (\ref{eq:zero}) in the case of flux line that follows 
\begin{widetext}
\begin{align}\label{eq:zform2}
u(r,\theta)\sim\frac{e^{-ikr}}{\sqrt{2\pi kr}}\sum_ {m=-\infty} ^ \infty c_m e^{im \theta}e^{i\varphi_{|m+\alpha|}} 
+\frac{e^{ikr}}{\sqrt{2\pi kr}}\sum_ {m=-\infty} ^ \infty c_m e^{im \theta}e^{-i\varphi_{|m+\alpha|}} \, . 
\end{align}
\end{widetext}
Therefore the equating the factor of incoming wave $ e^{-ikr}/\sqrt{r}$ gives the required $c_m$ to construct the eigenfunctions.
\begin{eqnarray}
c_m  = C i^m e^{-i\varphi_{|m+\alpha|}} \sum_ {n=-\infty} ^ \infty i^{n}\frac{\sin(\alpha-n)\pi }{(\alpha-n)\pi}e^{i\varphi_{m+n}}
 = C e^{-i\alpha \varepsilon}(-1)^{m+\alpha}(-i)^{|m+\alpha|}\, ,
\end{eqnarray}
where the series (\ref{eq:fourier}) is used to get the second line. Since the angle $\theta=-\pi$ is on the branch cut, the small positive value $0< \varepsilon\ll 1$ is added, e.i. $\theta=-\pi+\varepsilon$. At the final expression, it is understood that the limit $ \varepsilon \to 0$ should be considered eventually.

At this point we might check our $c_m$ by substituting it to the solution (\ref{eq:zero}) and take the vanishing flux limit $\alpha\to 0$. If everything is correct, final eigenfunction in this limit must be plane wave. That is, there is no scattering at all because of no flux and no hard cylinder as well. Indeed it is not hard to see $\sum c_m e^{im \theta} J_{| m+\alpha |}(kr)\to e^{ikr\cos \theta}$ as $\alpha \to 0$.

Comparing the coefficients of the out going radial cylindrical waves $e^{ikr}/\sqrt{r}$ in asymptotic forms (\ref{eq:abform1}) and (\ref{eq:zform2}) gives the scattering amplitude,
\begin{widetext}
\begin{align}\label{eq:zscatam}
f_{k, \varepsilon}(\alpha,\theta)&=\frac { e^{-i\alpha \frac{\varepsilon}{2}}}{\sqrt{2 i\pi k}}\sum_ {m=-\infty} ^ \infty e^{im \theta}[e^{-i\alpha \frac{\varepsilon}{2}} e^{i\pi(m+\alpha-|m+\alpha |)}-e^{i\alpha \frac{\varepsilon}{2}}]\nonumber\\
&=-\frac { e^{-i\alpha \frac{\varepsilon}{2}}}{\sqrt{2 i\pi k}}\sum_ {m > -\alpha} ^ \infty e^{im \theta}2i\sin(\alpha \varepsilon /2)+\frac { e^{-i\alpha \frac{\varepsilon}{2}}}{\sqrt{2 i\pi k}}\sum_ {-\infty} ^ {m<-\alpha} e^{im \theta} [e^{-i\alpha \frac{\varepsilon}{2}} e^{i2\pi(m+\alpha)}-e^{i\alpha \frac{\varepsilon}{2}}] 
\end{align}
\end{widetext} 
Here we have used the same $ \varepsilon$ trick to get a single summation by means of the expression (\ref{eq:fourier}). After some manipulations, $f_k(\alpha, \theta)$ becomes as

\begin{align}\label{eq:zscatam1}
f_{k, \varepsilon} (\alpha,\theta)=\frac { e^{-i\alpha \frac{\varepsilon}{2}}}{\sqrt{2 i\pi k}}\frac{\sin(\alpha \varepsilon /2)e^{-i([\alpha]+1/2) \theta}}{\sin(\theta /2)} 
+\frac { e^{i\alpha (\pi-\frac{\varepsilon}{2})}}{\sqrt{2 i\pi k}}\frac{\sin(\alpha (\pi-\varepsilon /2)e^{-i([\alpha]+1/2) \theta}}{\sin(\theta /2)}
\end{align}
where $[\alpha]$ is the integer part of the magnetic flux $\alpha$. There is no contribution that comes from $m+\alpha > 0$ as $ \varepsilon \to 0$. Therefore, the final expression for the scattering amplitude of the solenoid with zero radius becomes
\begin{equation}\label{eq:zscatam2}
f_{k} (\alpha,\theta)=\frac { \sin\pi\alpha}{\sqrt{2 i\pi k} }\frac{ e^{i\alpha \pi}e^{-i([\alpha]+1/2) \theta}}{\sin(\theta /2)}
\end{equation}
and the differential cross section is
\begin{equation}
\frac{d\sigma}{d \theta }=|f_{k} (\alpha,\theta)|^2=\frac { \sin^2\pi\alpha}{2 \pi k} \frac{ 1}{\sin^2(\theta /2)}\,,
\end{equation}
which is the AB original result apart from a phase factor \cite{AB59}.

Now it is possible to test if the scattering amplitude given in Eq.(\ref{eq:abscatam}) is reduced to the amplitude in (\ref{eq:zscatam2}) as the radius of the cylinder goes zero, $a\to 0$. After taking the limit $ \varepsilon \to 0$ and returning to the unscaled form of the amplitude  $f_k=\sqrt{a}\tilde{f}_{\tilde{k}}$, the summation can be divided in two parts by the conditions $m +\alpha > 0$ and $m +\alpha < 0$:
\begin{eqnarray}\label{eq:lscat}
-\sqrt{2\pi i k}f_k(\alpha,\theta)=\sum_ {m>-\alpha} ^ \infty e^{im \theta}\frac{2J_{m+\alpha}(ka)}{H_{m+\alpha}^{(1)}(ka)}
+\sum^{m<-\alpha}_{-\infty} e^{im \theta}\frac{2J_{m+\alpha}(ka)}{H_{m+\alpha}^{(1)}(ka)}
\end{eqnarray}
Using the limiting form of the Bessel and Hankel functions when $z\to 0$ and $\beta$ is fixed \cite{AS64},
\begin{eqnarray}
J_\beta(z)\sim \frac{(z/2)^\beta}{\Gamma(\beta+1)},\,\,\,\,(\beta\neq-1,-2,-3,\ldots)\\
H_\beta^{(1)}(z)\sim \frac{(z/2)^{-\beta}}{i\pi}\Gamma(\beta), \,\,\,\, ({\rm Re}(\beta) > 0)\, ,
\end{eqnarray}
it is easy to see that there is no contribution to the scattering amplitude from the first summation in Eq.(\ref{eq:lscat}). The above limit form of Hankel function can not be applied directly to the second summation because $m+\alpha < 0$ there. We must first apply the phase change for the Hankel function, $H_{-\beta}^{(1)}(z)=e^{\beta\pi i}H_\beta^{(1)}(z)$, and subsequently the limiting forms are applied in the second summation. It makes the summation independent of $ka$,
\begin{widetext}
\begin{align}
f_k(\alpha,\theta)=&\frac{2\pi i e^{-i([\alpha]+1)\theta}e^{i\nu\pi}}{-\sqrt{2\pi i k} }\sum_{m=0} ^\infty \frac{e^{-im \theta}e^{-i(m+1)\pi}}{\Gamma(1-(m+1-\nu))\Gamma(m+1-\nu)}=\frac { \sin\pi\nu}{\sqrt{2 \pi i k} }\frac{ e^{i\nu \pi}e^{-i([\alpha]+1/2) \theta}}{\sin(\theta /2)}\ .
\end{align}
\end{widetext}
Here we use $\Gamma(1-z)\Gamma(z)=\pi/\sin\pi z$ to obtain the last expression, which is identical to Eq.(\ref{eq:zscatam2}). Here $\nu$ is the decimal part of the flux $\alpha$.

\section{Conclusion and discussion} \label{sec:conclusion}
An unbound eigenvalue equation has been solved as a scattering problem. A charged particle is scattered from a impenetrable cylinder enclosing a solenoid as an electromagnetic source. The magnetic induction field ${\bf B}$ inside the solenoid can not prevent the particle from entering the cylinder region. Therefore, it is supposed an infinite scalar potential barrier  
exists in the circular region in additional to the magnetic vector potential.

The scattering amplitude and corresponding total cross section of a charged particle from a solenoid with finite radius are obtained as a sum over angular momentum quantum number. Different limiting cases are checked to test the result. The scattering amplitude must reduce to the hard cylinder scattering case when the radius of the cylinder is kept constant and the magnetic flux goes to zero. It must end up with the AB formula when the flux is fixed and the radius goes to zero. When both the flux and the radius vanish, it must give zero amplitude, i.e., no scattering at all. That is, the solution is just free particle in empty space. The result of this work fulfils all these requirements. Only negative angular momentum eigenvalues $m+\alpha<0$ contribute to the AB scattering amplitude ($a=0$) while all eigenvalues play the role for the amplitude of the solenoid with finite radius, $a\neq 0$ (see Eqs. (\ref{eq:zscatam}) and (\ref{eq:lscat})) .

The charged particle does not touch the source and its magnetic induction field inside the solenoid, thus the classical dynamics of the particle is the same with or without magnetic flux. But the flux still can create a measurable effect in the quantum level if the flux $\alpha$ is a non-integer real number (see Eq. (\ref{eq:abtcs})). It is worth mention that kinematical linear momentum or angular momentum is a gauge invariant quantity while the canonical ones are not. However, the quantization is performed by the canonical momenta and their conjugate coordinates. That is why AB effect is a purely quantum mechanical phenomena. The vector potential is just part of the kinematic momentum or the flux is the some part of the angular momentum around the symmetry axis (z$-$axis).

In field theoretical point of view, it is important to solve Dirac's equation in the same way like this work to include the relativistic spin effect for the AB scattering explicitly. It is also worth to study the time dependent scattering by considering the time varying current on the solenoid (Faraday induction and retardation effect). These are the future considerations at the moment. Moreover, less obvious feature of the problem is the following: the simple loop integration around the singularity is independent the shape of the loop so that it is called a topological property of the space (doubly connected region in the present case). The singularity in the forbidden region might be something else instead of the solenoid, such as electric charge, magnetic monopole, or black hole singularity. It might also be interesting to investigate those.  
 
\begin{acknowledgments}
The author is grateful to C. Harabati and to O. Tosun for helpful discussions. 
\end{acknowledgments}

%----------------------------------------------------------------------
%\begin{figure*}
%\centering
%\epsfig{figure=f1.eps,scale=0.7}
%\epsfig{figure=f2.pdf,scale=0.6}
%\caption{Recommended positron binding energies from Table \ref{t:5} to the %dissociation threshold e$^++$A. The results based on current study are } 
%\label{f:2}
%\end{figure*}
%----------------------------------------------------------------------

\end{document}